\documentclass[reprint,twocolumn,superscriptaddress,showpacs,nofootinbib,notitlepage,preprintnumbers]{revtex4-2}
\usepackage{CJK} 
\usepackage{graphicx}
\usepackage{hyperref}
\usepackage[dvipsnames]{xcolor}
\usepackage{amsmath,amsthm,amssymb}
\usepackage{comment}
\usepackage{enumerate}
\usepackage{lipsum}

\allowdisplaybreaks[4]

\begin{document}

\title{Kolmogorov-type non-thermal fixed points and beyond of far-from-equilibrium dilute system: ultra-cold Fermi gas}

\author{Chun-Wei Su}
\email{chsu1718@colorado.edu}
\affiliation{Department of Physics, University of Colorado, Boulder, Colorado 80309, USA}

\date{\today}

\begin{abstract}
The far-from-equilibrium dynamics driven by the scattering from next-to-leading-order (NLO) corrections in the quantum field theory has stationary solutions for the particle distribution characterized as the Kolmogorov-type non-thermal fixed points corresponding to scale-invariant particle/energy flux. The dynamics of the spatially homogeneous dilute ultra-cold Fermi gas is investigated, and its kinetic equation confirms the Kolmogorov-type non-thermal fixed points in the perturbation theory by the quasi-particle assumption, in contrast to the wave turbulence of the ultra-cold Bose gas. In addition, other stationary states are discussed when the the vertex correction becomes significant in the kinetic equation. It is found that conditions for the scale-invariant flux in this case no longer hold when the quasi-particle assumption is applied.
\end{abstract}

\maketitle

\section{Introduction}\label{sec:introduction}
The Kolmogorov-type non-thermal fixed point is the non-equilibrium stationary solution of the kinetic equation \cite{Zakharov:2013ko} originating from the study of hydrodynamic turbulence in the incompressible fluid by Kolmogorov \cite{Kolmogorov:1941}. It features power-law spectra and distribution. Analogously, other systems not restricted to the hydrodynamic one also show power-law behavior first found by Zakharov \cite{1965JAMTP} in weakly interacting waves, called wave turbulence or Kolmogorov-Zakharov turbulence \cite{Zakharov:1992za}, which is easier than the hydrodynamic turbulence to access by solving the wave kinetic equation. The power-law solutions of the kinetic equation describe the stationary transport in the presence of external source and sink that ensure constant flux.

These non-thermal states are not restricted to the classical systems since the kinetic description can also be extended to the non-equilibrium quantum field theory used in the evolution of the early universe \cite{Micha:2002ey,Berges:2002cz,Micha_2004,Gasenzer_2012} and the formation of Bose-Einstein condensation \cite{Gasenzer_2009,Berges_2012,Nowak_2014,PineiroOrioli:2015cpb}. The field-theoretic approach builds up the time evolution equations of two-point functions, called the Kadanoff-Baym equations \cite{kadanoff1962quantum,kadanoff1962quantum,PhysRevD.37.2878,Berges_2006} by including scatterings driven by quantum corrections. The kinetic equation and beyond can be derived by the gradient expansion of the Kadanoff-Baym equations at late time, when solutions of real-time evolution can be analytically evaluated for a homogeneous system.

Once the particle number distribution function $f(t,\textbf{p})$ satisfies $f(t,\textbf{p})\gg1$ for a Bosonic system, the kinetic theory is described by the wave kinetic equation \cite{Zakharov:1992za}, which is usually constructed by the perturbation theory, except in the over-occupied situation $\lambda^{-1}\gtrsim f(t,\textbf{p})\gg1$ where $\lambda$ is the coupling. In this situation, the kinetic theory breaks down since a systematic weak-coupling expansion is spoiled. Higher-order corrections can dominate the collision kernel. Therefore, the weakly coupled system with highly occupied states becomes strongly correlated by the significant nonlinear effect. The generalization to the vertex-resummed kinetic theory \cite{Walz_2018} fixes the issue by replacing the expansion parameter with the inverse number of field components $1/N$ \cite{PhysRevD.10.2428,Berges_2002,Aarts:2002dj}. The resummed loop diagrams in the vertex add corrections to the coupling in the IR regime, which shows different scaling solutions than the perturbative result \cite{Berges_2008,Berges:2010ez,Walz_2018}.

Based on the field-theoretic approach for the non-equilibrium quantum system, non-thermal stationary solutions obtained by solving the kinetic equation are characterized by the power-law distribution in terms of momenta $|\textbf{p}|$ related to scale-invariant fluxes like what classical systems show up. The Kolmogorov-Zakharov turbulence is theoretically and numerically confirmed in perturbation theory and also the strong wave turbulence, in contrast to the weak wave turbulence, is found tp describe the strongly correlated system realized by over-occupied states in the IR regime \cite{Berges_2008,Berges:2010ez}.

The kinetic theory of the dilute ultra-cold Fermi gas in the limit $f(t,\textbf{p})\ll1$, on the other hand, can be described by the Boltzmann equation, which accounts for collisions between particles in contrast to waves in the bosonic system in the limit $f(t,\textbf{p})\gg 1$. The exact stationary solutions characterized as Kolmogorov-type non-thermal fixed points for the Boltzmann equation were first found by Kats, Kontorovich, Moiseev and Novikov \cite{PismaZhETF.21.13,Kats:1976}.

In this paper, the non-equilibrium dynamics is studied in the spatially homogeneous dilute ultra-cold Fermi gas by including NLO corrections as scattering effects in the quantum field theory. The time evolution equations of two-point functions are derived, which become the kinetic equation at the late time after performing and keeping the leading term in the gradient expansion. By solving the kinetic equation, Kolmogorov-type non-thermal solutions are found related to the particle and energy conservation \cite{PismaZhETF.21.13,Kats:1976} relying on the quasi-particle assumption (width is vanishing) in the perturbation theory, which is different from the weak wave turbulence in Ref.~\cite{Scheppach:2009wu}.

Solutions beyond Kolmogorov-type non-thermal fixed points are discussed in the situation that the vertex correction is large and requires a resummation performed in a systematic expansion based on the small parameter $1/N$ in the large $N$ limit \cite{rosenhaus2024strongwaveturbulencestrongly,Romatschke:2019rjk,Romatschke:2019wxc}. A distinct scaling behavior added by the resummation will modify the scaling structure of kinetic equation and the scaling laws related to the scale-invariant flux.



\section{Weak-coupling and $1/N$ expansion}

The ultra-cold Fermi gas in the quantum field theory is described by fermionic field operators $\hat{\psi}_{\alpha }^i$, $\hat{\psi}_{\alpha }^{i\dagger}$ with corresponding Grassmann variables $\psi_\alpha^i$, $\psi_\alpha^{i\dagger}$ labeled by the spin ($\alpha=\uparrow,\downarrow$) and the field's component ($N$ components in total) as the lower and the upper indices, respectively. Grassmann variables $\psi_\alpha^i$ and $\psi_\alpha^{i\dagger}$ can be re-formulated by the real part $\psi_{\alpha 1}^i$ and the imaginary part $\psi_{\alpha 2}^i$
\begin{align}
&\,\psi_\alpha^i=\frac{1}{\sqrt{2}}(\psi_{\alpha 1}^i+i\psi_{\alpha 2}^i)\,,\,\, \quad \psi_\alpha^{i\dagger}=\frac{1}{\sqrt{2}}(\psi_{\alpha 1}^i-i\psi_{\alpha 2}^i).
\end{align}
In addition, I define $\bar{\psi}_{\alpha j}^i\equiv-\psi_{\beta k}^i\,\mathcal{I}_{\beta\alpha}\otimes(\sigma_y)_{kj}$ \cite{Kronenwett:2010ic,kronenwett2010nonthermalequilibrationonedimensionalfermi}, where $\mathcal{I}$ is the identity matrix and $\sigma_y$ is the second Pauli matrix; repeated indices for spin, field's component and real/imaginary component imply summation, unless the indices are primed. To clarify the notation in the following, spin is always denoted by Greek letters; field's component and real/imaginary component are represented by Latin letters.

The action of the ultra-cold Fermi gas is
\begin{align}
S[\psi_\alpha^{i\dagger},\psi_\alpha^i]=&\,\int_{x,\mathcal{C}}\mathcal{L}\,,\\
\mathcal{L}=&\,\psi_\alpha^{i\dagger}(x)\bigg(i\partial_{x^0}+\frac{\nabla_\textbf{x}^2}{2m}\bigg)\psi_\alpha^i(x)\nonumber\\
&\,-\frac{4\pi a_s}{Nm}\psi_\uparrow^{k\dagger}(x)\psi_\downarrow^{l\dagger}(x)\psi_\downarrow^l(x)\psi_\uparrow^k(x),
\end{align}
where $a_s$ is the s-wave scattering length and $m$ denotes the fermion mass; $\int_{x,\mathcal{C}}\equiv\int_{-\infty}^\infty d^{d-1}x\int_\mathcal{C} dx^0$ is defined with $d$ as the space-time dimension. The time integral is performed on the Schwinger-Keldysh contour $\mathcal{C}$ \cite{Schwinger:1960qe,Keldysh:1964ud} going to $t=\infty$ from an initial time $t^0$ and back to $t^0$ along the real-time axis and the path ordering is obeyed. Therefore, the Dirac delta function $\delta_{ \mathcal{C} } (x-y)\equiv\delta_\mathcal{C}(x^0-y^0)\delta^{d-1}(\textbf{x}-\textbf{y})$ defined on the closed time contour $\mathcal{C}$ is zero everywhere unless both two point $x$ and $y$ satisfy $x=y$ on either $\mathcal{C}^+$ (forward path) or $\mathcal{C}^-$ (backward path) together with
\begin{align}
\int_\mathcal{C}dx^0f(x^0)\,\delta_\mathcal{C}(x^0-y^0)=f(y^0),
\end{align}
which implies that $\delta_\mathcal{C}(x^0-y^0)=\pm\delta(x^0-y^0)$ when $y^0$ is on $\mathcal{C}^\pm$. Similarly, the step function $\theta_\mathcal{C}(x^0-y^0)$ and ${\rm sgn}_\mathcal{C}(x^0-y^0)\equiv\theta_\mathcal{C}(x^0-y^0)-\theta_\mathcal{C}(y^0-x^0)$ are defined by the path ordering on $\mathcal{C}$ as well.

The action can be re-written as
\begin{align}
S[\psi_{\alpha j}^i]
=&\,\,\int_{x,\mathcal{C}}\bigg[\frac{1}{2}\bigg(\bar{\psi}_{\alpha j}^i(x)\Big(-\mathcal{I}_{\alpha\beta}\otimes (\sigma_y)_{jk}\Big)i\partial_{x^0}\psi_{\beta k}^i(x)\nonumber\\
&\,+\bar{\psi}_{\alpha j}^i(x)\Big(\mathcal{I}_{\alpha\beta}\otimes\mathcal{I}_{jk}\Big)\frac{\nabla_\textbf{x}^2}{2m}\psi_{\beta k}^i(x)\bigg)\nonumber\\
&\,-\frac{\pi a_s}{Nm}\bar{\psi}_{\uparrow j}^i(x)\psi_{\uparrow j}^i(x)\bar{\psi}_{\downarrow k}^l(x)\psi_{\downarrow k}^l(x)\bigg],
\label{eq:acSS}
\end{align}
by the relation $\psi_{\alpha 1}^i(x)\psi_{\alpha 1}^i(x)=\psi_{\alpha 2}^i(x)\psi_{\alpha 2}^i(x)=0$ used in quartic terms.

The time-ordered two-point function $G_{\alpha\beta,ij}^{kl}(x,y)\equiv \langle\mathcal{T}_\mathcal{C}\, \hat{\psi}_{\alpha i}^k(x)\hat{\bar{\psi}}_{\beta j}^l(y)\rangle$ for the fermion can be constructed with $\hat{\bar{\psi}}_{\beta j}^l\equiv-\hat{\psi}_{\alpha k}^l\mathcal{I}_{\alpha\beta}\otimes(\sigma_y)_{kj}$ and $\mathcal{T}_\mathcal{C}$ as the time ordering on $\mathcal{C}$. By the $O(N)$ symmetry of the system in the space of field's component in Eq.~(\ref{eq:acSS}), the time-ordered two-point function is diagonalized as $G_{\alpha\beta,ij}^{kl}(x,y)=G_{\alpha\beta,ij}(x,y)\delta^{kl}$ with $G_{\alpha\beta,ij}(x,y)\equiv \langle\mathcal{T}_\mathcal{C}\, \hat{\psi}_{\alpha i}(x)\hat{\bar{\psi}}_{\beta j}(y)\rangle$, which can be decomposed into the statistical $F_{\alpha\beta,ij}$ and the spectral function $\rho_{\alpha\beta,ij}$ defined by
\begin{align}
G_{\alpha\beta,ij}(x,y)=&\,F_{_{\alpha\beta,ij}}(x,y)
-\frac{i}{2}\rho_{_{\alpha\beta,ij}}(x,y)\nonumber\\
&\,\times{\rm sgn}_\mathcal{C}(x^0-y^0)\,, \label{eq:rrnG}\\
F_{_{\alpha\beta,ij}}(x,y)\equiv&\,\frac{1}{2}\big\langle\big [ \hat{\psi}_{\alpha i}(x),\hat{\bar{\psi}}_{\beta j}(y)\big]\big\rangle ,\\
\rho_{_{\alpha\beta,ij}}(x,y)\equiv&\,i \big\langle \big\{ \hat{\psi}_{\alpha i}(x),\hat{\bar{\psi}}_{\beta j}(y) \big\}\big\rangle\,,
\end{align}
which have the symmetry property
\begin{align}
F_{_{\alpha\beta,ij}}(x,y)&\,
=(-1)^{i^\prime+j^\prime}F_{_{\beta\alpha,(3-j^\prime)(3-i^\prime)}}(y,x)\,, \label{eq:FRRR}\\
\rho_{_{\alpha\beta,ij}}(x,y)&\,=(-1)^{i^\prime+j^\prime+1}\rho_{_{\beta\alpha,(3-j^\prime)(3-i^\prime)}}(y,x).\label{eq:rrgg}
\end{align}

The quantum corrections to the two-point function are derived through the resummation technique shown in Ref.~\cite{Romatschke:2019rjk,Romatschke:2019wxc}. People usually applied the two-particle irreducible (2PI) eﬀective action approach \cite{Berges_2002,Aarts:2002dj,Kronenwett:2010ic} instead, two methods achieve identical results though. That is, the NLO contributions to the two-point function by the 2PI correspond to corrections characterized as Resummation Level Two (R2) \cite{Romatschke:2019rjk} in this study. It will be seen that the NLO corrections in the weak-coupling or $1/N$ expansion lead to scattering processes beyond the mean field.

\subsection{Weak-coupling expansion: Leading order (LO)} \label{sec:WeLO}
In order to capture the LO and NLO corrections to the self-energies, the R2-resummation is constructed by the re-organization of the action as below 
\begin{align}
iS=&\,iS_0+iS_I=iS_0^{\rm R2}+iS_I^{\rm R2}, \\
iS_0^{\rm R2}=&\,-\frac{1}{2}\int_{x,\mathcal{C}}\int_{y,\mathcal{C}}\bar{\psi}_{\alpha i}^l(x)G_{\alpha\beta,ij}^{-1}(x,y)\psi_{\beta j}^l(y),\\
iS_I^{\rm R2}=&\,\frac{1}{2}\int_{x,\mathcal{C}}\int_{y,\mathcal{C}}\bar{\psi}_{\alpha i}^l(x)\Sigma_{\alpha\beta,ij}(x,y)\psi_{\beta j}^l(y)\nonumber\\
&\,-i\frac{\pi a_s}{Nm}\int_{x,\mathcal{C}}\int_{y,\mathcal{C}}\bar{\psi}_{\uparrow i}^l(x)\psi_{\uparrow i}^l(x)\bar{\psi}_{\downarrow j}^r(x)\psi_{\downarrow j}^r(x)\,,
\end{align}
where
\begin{align}
G_{0_{\alpha\beta,ij}}^{-1}(x,y)\equiv&\,-i\Big(-\mathcal{I}_{\alpha\beta}\otimes (\sigma_y)_{ij}\Big)\delta_{\mathcal{C}}(x-y)\,i\partial_{x^0} \nonumber\\
&\, -i\,\mathcal{I}_{\alpha\beta}\otimes \mathcal{I}_{ij}\delta_{\mathcal{C}}(x-y)\,\frac{\nabla_\textbf{x}^2}{2m}
\label{eq:Gfre}
\end{align}
is the inverse of time-ordered two-point function for the free fermion and
\begin{align}
G_{\alpha\beta,ij}^{-1}(x,y)\equiv&\,G_{0_{\alpha\beta,ij}}^{-1}(x,y)+\Sigma_{\alpha\beta,ij}(x,y).
\label{eq:Ginv}
\end{align}
So far, the two-point function $G_{0_{\alpha\beta,ij}}$ has been promoted to $G_{\alpha\beta,ij}$ by including undetermined self-energies $\Sigma_{\alpha\beta,ij}$. In addition, any additional term present in $iS_0^{\rm R2}$ is subtracted in $iS_I^{\rm R2}$ in order to maintain the expression of the action.
The local contributions at the NLO order to $\Sigma_{\alpha\beta,ij}$ are determined
\begin{align}
\Sigma_{\alpha\beta,ij}(x,x)=
\begin{cases}
\quad0 \quad\quad\quad\quad\quad\quad\,\,\,\,\,\,\quad\,\,\,\, \alpha\neq\beta \\\\
        -i\frac{2\pi a_s}{m}G_{\bar{\alpha}\bar{\beta},kk}(x,x)\delta_{ij}\,\,\,\,\alpha=\beta\\
\end{cases} \label{eq:S0uu}
& {{\rm at} \,\,{\rm LO}}\,,
\end{align}
where $\bar{\uparrow}\equiv\,\downarrow (\bar{\downarrow}\equiv\,\uparrow)$, by evaluating the two-point function of the fermion to the order linear in $iS_I^{\rm R2}$. The corresponding Feynman diagram is shown in Fig.~\ref{fig:Feth}, which is vanishing since the spin-diagonal state is considered.

\subsection{Weak-coupling expansion: Beyond LO } \label{sec:WeNLO}
\begin{figure}[phtb]
\begin{center}
\includegraphics[scale=0.25]{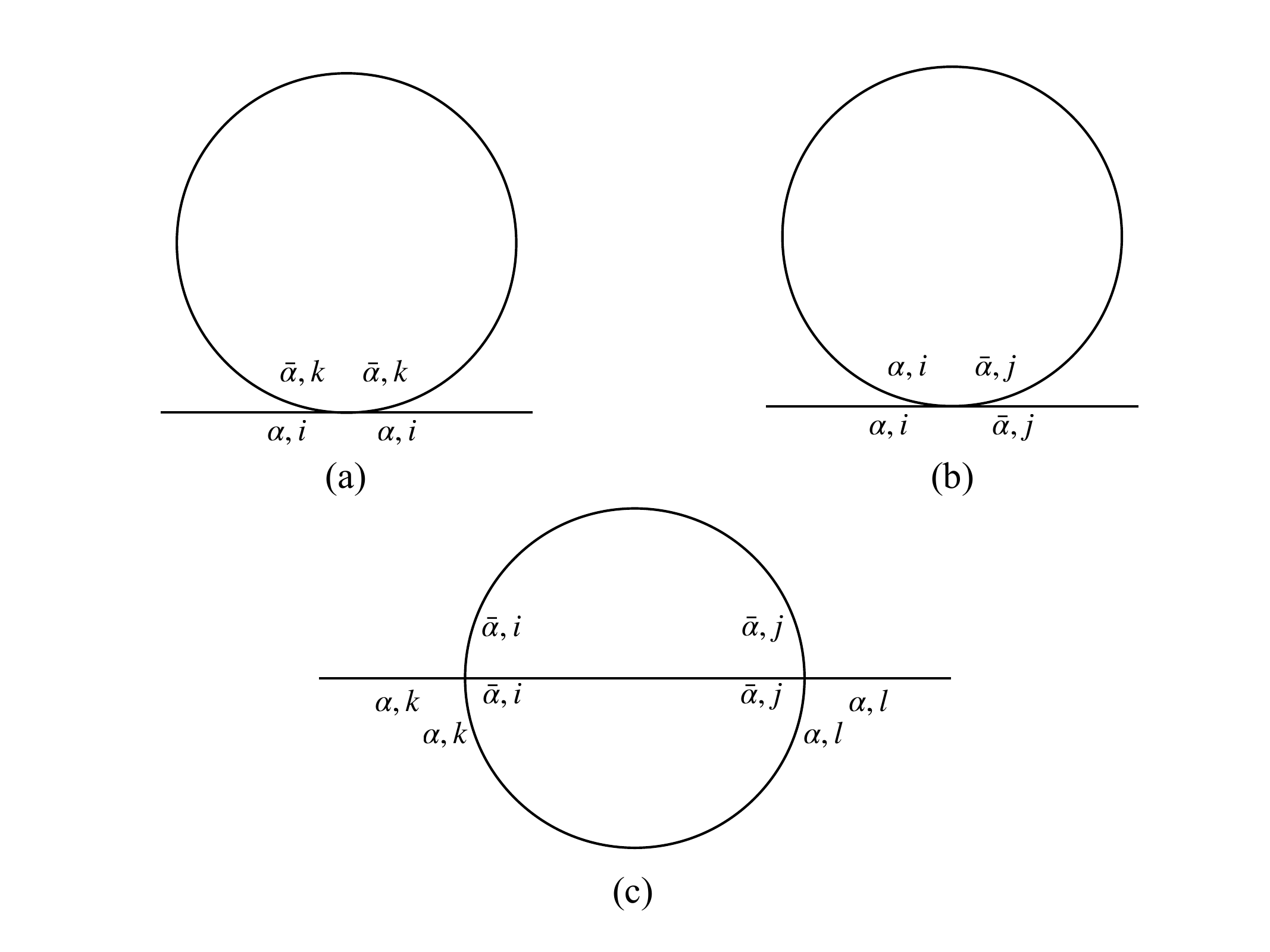}
 \caption{Feynman diagrams for self-energies shown in Eq.~(\ref{eq:S0uu}) in (a), Eq.~(\ref{eq:S0uN}) in (b) and Eq.~(\ref{eq:suns}) in (c). The solid line denotes the fermion's two-point function.} 
\label{fig:Feth}
\end{center}
\end{figure}

Corrections beyond LO to $\Sigma_{\alpha\beta,ij}$ have local and non-local contributions. Along with Sec.~\ref{sec:WeLO} to the order linear in $iS_I^{\rm R2}$, local corrections at the NLO are obtained as
\begin{align}
\Sigma_{\alpha\beta,ij}(x,x)=
\begin{cases}
\frac{i}{N}\frac{4\pi a_s}{m}G_{\alpha\beta,ij}(x,x) \quad\alpha\neq\beta \\\\
        \quad0\quad \quad\quad\quad\quad\quad\quad\,\,\,\alpha=\beta\\
\end{cases} \label{eq:S0uN}
{\rm at\,\,\,NLO},
\end{align}
expressed as the Feynman diagram in Fig.~\ref{fig:Feth}.

Following a similar procedure for the LO term, non-local self-energies are specified at the order of $\big(iS_I^{\rm R2}\big)^2$ by including only the first order of $\Sigma_{\alpha\beta,ij}(x,y)$. The resulting non-trivial self energies are
\begin{align}
&\,\Sigma_{\alpha^\prime\alpha^\prime,kl}(x,y) \nonumber\\
=
&\,-\frac{8}{N}\bigg(\frac{\pi a_s}{m}\bigg)^2G_{\alpha^\prime\alpha^\prime,kl}(x,y)G_{\bar{\alpha}^\prime\bar{\alpha}^\prime,ij}(x,y)G_{\bar{\alpha}^\prime\bar{\alpha}^\prime,ji}(y,x).
\label{eq:suns}
\end{align}
The Feynman diagram of Eq.~(\ref{eq:suns}) is visualized in Fig.~\ref{fig:Feth}

Note that any off-diagonal element of the self-energy $\Sigma_{\uparrow\downarrow(\downarrow\uparrow),ij}(x,y)$ in the spin space has at least a factor of $G_{\uparrow\downarrow(\downarrow\uparrow),ij}(x,y)$. Therefore, non-local self-energy $\Sigma_{\uparrow\downarrow(\downarrow\uparrow),ij}(x,y)$ is absent at the NLO since $G_{\uparrow\downarrow(\downarrow\uparrow),ij}(x,y)$, compared with $G_{\uparrow\uparrow(\downarrow\downarrow),ij}(x,y)$, is beyond the leading contribution. Similarly, the diagonal element $\Sigma_{\uparrow\uparrow(\downarrow\downarrow),ij}(x,y)$ expanded up to two loops has omitted any term with $G_{\uparrow\downarrow(\downarrow\uparrow),ij}(x,y)$.

\subsection{\texorpdfstring{$1/N$}{1/N} expansion: LO} \label{sec:LNLO}
The LO contributions in the $1/N$ expansion can be easily obtained by performing the Hubbard-Stratonovich transformation, by which any diagrams contributing to the LO mean-field approximation are resummed. In this case, the transformation introduces four auxiliary fields to construct an identity 
\begin{align}
&\,\int\mathcal{D}\mathcal{\sigma}\mathcal{D}\mathcal{\bar{\sigma}}\mathcal{D}\zeta\mathcal{D}\bar{\zeta}\, {\rm exp}\bigg[i\int_{x,\mathcal{C}}\bigg(\bar{\zeta}(x)\Big(\bar{\sigma}(x)-\bar{\psi}_{\uparrow j}^i(x)\psi_{\uparrow j}^i(x)\Big) 
\nonumber\\
&\,+\zeta(x)\Big(\sigma(x)-\bar{\psi}_{\downarrow j}^i(x)\psi_{\downarrow j}^i(x)\Big)\bigg)\bigg]\,,
\end{align}
which multiplies the path integral so that the generating functional becomes
\begin{align}
Z=&\int \mathcal{D}\psi_\alpha^i\mathcal{D}\bar{\zeta}\mathcal{D}\zeta\nonumber\\
&\,\,{\rm exp}\bigg\{\int_{x,\mathcal{C}}\bigg[-\frac{1}{2}\bar{\psi}_{\alpha i}^l(x)G_{0_{\alpha\beta,ij}}^{-1}(x,y)\psi_{\beta j}^l(x)\nonumber\\
&\,-i\bar{\zeta}(x)\bar{\psi}_{\uparrow j}^l(x)\psi_{\uparrow j}^l(x)-i\zeta(x)\bar{\psi}_{\downarrow i}^l(x)\psi_{\downarrow i}^l(x)\nonumber\\
&\,+i\frac{Nm}{\pi a_s}\bar{\zeta}(x)\zeta(x)\bigg]\bigg\}
\end{align}
after integrating out auxiliary fields $\sigma(x)$ and $\bar{\sigma}(x)$. By splitting auxiliary fields $i\zeta(x)\equiv i\Delta(x)+i\zeta^\prime(x)$ and $i\bar{\zeta}(x)\equiv i\bar{\Delta}(x)+i\bar{\zeta}^\prime(x)$ into zero modes $\Delta(x),\bar{\Delta}(x)$ and fluctuations $\zeta^\prime(x),\bar{\zeta}^\prime(x)$, the LO generating functional $Z_{\rm R0}$ identified as the R0-level contribution in Ref.~\cite{Romatschke:2019rjk} is given by
\begin{align}
Z_{\rm R0}=&\int \mathcal{D}\psi_\alpha^i \mathcal{D}\bar{\zeta}^\prime\mathcal{D}\zeta^\prime d\bar{\Delta}d\Delta\nonumber\\
&\,\,{\rm exp}\bigg\{\int_{x,\mathcal{C}}\bigg[-\frac{1}{2}\bar{\psi}_{\alpha i}^l(x)G_{0_{\alpha\beta,ij}}^{-1}(x,y)\psi_{\beta j}^l(x)\nonumber\\
&\,-i\bar{\Delta}(x)\bar{\psi}_{\uparrow j}^l(x)\psi_{\uparrow j}^l(x)-i\Delta(x)\bar{\psi}_{\downarrow i}^l(x)\psi_{\downarrow i}^l(x)\nonumber\\
&\,+i\frac{Nm}{\pi a_s}\bar{\Delta}(x)\Delta(x)+i\frac{Nm}{\pi a_s}\bar{\zeta}^\prime(x)\zeta^\prime(x)\bigg]\bigg\}.
\end{align}

\subsection{\texorpdfstring{$1/N$}{1/N} expansion: Beyond LO} \label{sec:LNNLO}

\begin{figure}[phtb]
\begin{center}
\includegraphics[scale=0.3]{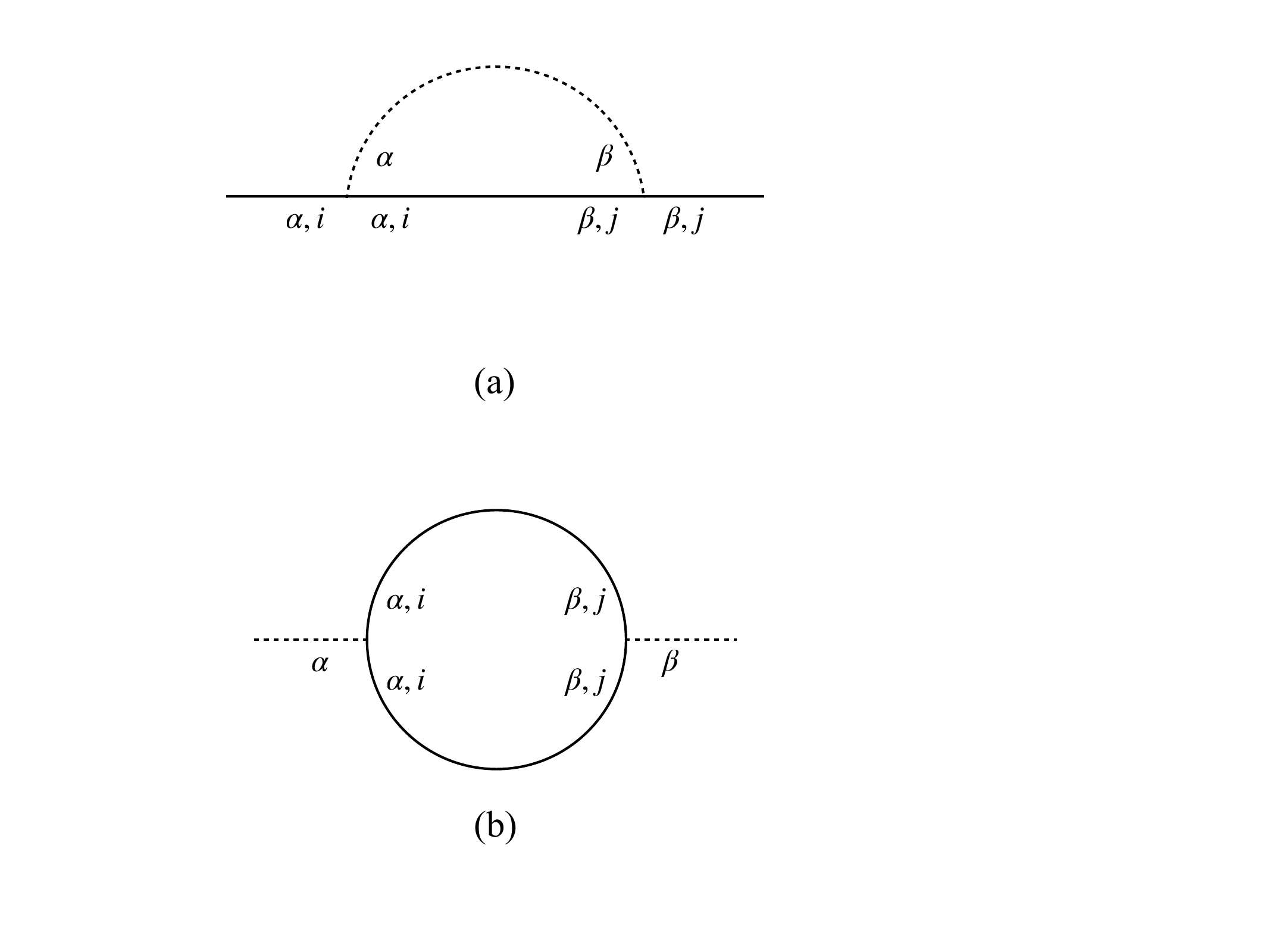}
 \caption{Feynman diagrams of self-energies shown in Eq.~(\ref{eq:SSuu}) in (a) and Eq.~(\ref{eq:Piuu}) in (b). The dashed line is the auxiliary field's two-point function.}
\label{fig:Fetw}
\end{center}
\end{figure}

Compared with the NLO weak-coupling expansion shown in Sec.~(\ref{sec:WeNLO}), the procedure of the R2 resummation in the $1/N$ expansion is similar except for the presence of the auxiliary fields' two-point function $D_{\alpha\beta}(x,y)$ and their self-energies $\Pi_{\alpha\beta}(x,y)$
\begin{align} 
iS=&\,iS_0+iS_I=i\int_{x,\mathcal{C}}\frac{Nm}{\pi a_s}\bar{\Delta}\Delta+iS_0^{\rm R2}+iS_{I}^{\rm R2}, \label{eq:SoR2}\\
iS_0^{\rm R2}=&\, -\frac{1}{2}\int_{x,\mathcal{C}}\int_{y,\mathcal{C}}\, \bar{\psi}_{\alpha i}^l(x) G_{\alpha \beta , i j}^{-1}(x,y)\psi_{\beta j}^l(y)\nonumber\\
&\,-\frac{m}{\pi}\int_{x,\mathcal{C}}\int_{y,\mathcal{C}} \,\frac{1}{2}\,\chi^\intercal(x)D^{-1}(x,y)\chi(y),\label{eq:SIR1}\\
iS_I^{\rm R2}=&\,\frac{1}{2} \int_{x,\mathcal{C}}\int_{y,\mathcal{C}}\, \bar{\psi}_{\alpha i}^l(x) \Sigma_{\alpha\beta, ij}(x,y)\psi_{\beta j}^l(y)\nonumber\\
&\,-\frac{m}{\pi} \int_{x,\mathcal{C}}\int_{y,\mathcal{C}}\,\frac{1}{2}\,\chi^\intercal(x)N\Pi(x,y)\chi(y)\nonumber\\
&\,-\int_{x,\mathcal{C}}i\bar{\zeta}^\prime(x)\bar{\psi}_{\uparrow j}^l(x) \psi_{\uparrow j}^l(x)\nonumber\\
&\,-\int_{x,\mathcal{C}}i\zeta^\prime (x)\bar{\psi}_{\downarrow i}^l(x)\psi_{\downarrow i}^l(x)\,,
\label{eq:SIR2}
\end{align}
where
\begin{align}
D_{\alpha\beta}^{-1}(x,y)=&\,iN\bigg(-\frac{(\sigma_x)_{\alpha\beta}}{a_s}\delta_\mathcal{C}(x-y)+i\Pi_{\alpha\beta}(x,y)\bigg), \\
\label{eq:Dmox}
\chi(x)=&\,
\begin{pmatrix}
\bar{\zeta}^\prime(x) \\
\zeta^{\prime}(x)
\end{pmatrix}
\,,
\end{align}
and $\sigma_x$ is the first Pauli matrix. The resulting self-energies of the fermion, shown in Fig.~\ref{fig:Fetw}, are 
\begin{align}
\Sigma_{\alpha\beta,ij}(x,y)=&\,\frac{4\pi}{m}D_{\alpha^\prime\beta^\prime}(x,y)G_{\alpha^\prime\beta^\prime,ij}(x,y)\label{eq:SSuu}
\end{align}
obtained by calculating the two-point function of the fermion field to the order $\big(iS_I^{\rm R2}\big)^2$ with an order of $\Sigma_{\alpha\beta,ij}(x,y)$. In fact, the NLO mean-field contributions can be be derived by reducing the R2 to the R1 resummation by $\Pi(x,y)=0$ \cite{Romatschke:2019rjk} such that $D_{\alpha\beta}(x,y)=i\frac{a_s}{N}(\sigma_x)_{\alpha\beta}\delta_{\mathcal{C}}(x-y)$. Therefore, I get
\begin{align}
\Sigma_{\alpha\beta,ij}(x,x)=
\begin{cases}
\frac{i}{N}\frac{4\pi a_s}{m}G_{\alpha\beta,ij}(x,x) \quad\,\,\,\alpha\neq\beta \\\\
        \quad0\quad \quad\quad\quad\quad\quad\quad\quad\alpha=\beta\\
\end{cases} 
&{\rm at}\,\, \,{\rm NLO},
\end{align}
which is the same as Eq.~(\ref{eq:S0uN}).

On the other hand, self-energies $\Pi_{\mu\nu}(x,y)$ are obtained by calculating the two-point function of the auxiliary fields to the order $\big(iS_I^{\rm R2}\big)^2$ with an order of $\Pi_{\alpha\beta}(x,y)$
\begin{align}
\Pi_{\alpha\beta}(x,y)=&\,\frac{2\pi}{m}G_{\alpha^\prime\beta^\prime,ij}(x,y)G_{\beta^\prime\alpha^\prime,ji}(y,x)\,,\label{eq:Piuu}
\end{align}
which is shown in Fig.~\ref{fig:Fetw}. It is noticed that non-local off-diagonal elements of self-energies in the spin space are sub-leading since $G_{0_{\uparrow\downarrow(\downarrow\uparrow),ij}}(x,y)$ is beyond the leading contribution, so $\Pi_{\uparrow\downarrow(\downarrow\uparrow)}(x,y)$ can be ignored in the following discussions.

Finally, the corresponding effective action $\Gamma_{\rm eff}^{\rm R2}$ per component at the R2 level is constructed by Eq.~(\ref{eq:SoR2}), (\ref{eq:SIR1}) and (\ref{eq:SIR2})
\begin{align}
\Gamma_{\rm eff}^{\rm R2}
=&\,i\frac{m}{\pi a_s}{\rm Tr} \big[\bar{\Delta}(x)\Delta(x)\big]+\frac{1}{2}{\rm Tr}\ln G^{-1}(x,y)\nonumber\\
&\,-\frac{1}{2N}{\rm Tr}\ln\Big[\frac{m}{\pi}D^{-1}(x,y)\Big]-\frac{1}{2}{\rm Tr}\big[\Sigma(x,y)G(y,x)\big],
\end{align}
where the trace is applied to sum over the space-time and all the other indices. With the help of Eq.~(\ref{eq:SSuu}) and (\ref{eq:Piuu}), zero modes $\Delta(x)$ and $\bar{\Delta}(x)$ are determined by saddle-point conditions
\begin{align}
\frac{\delta \Gamma_{\rm eff}^{\rm R2}}{\delta \bar{\Delta}(x)}=&\,0\quad \Rightarrow \quad \Delta(x)=-\frac{\pi a_s}{m}G_{\uparrow\uparrow,ii}(x,x)\,,\label{eq:LNlu}\\
\frac{\delta \Gamma_{\rm eff}^{\rm R2}}{\delta \Delta(x)}=&\,0\quad \Rightarrow \quad \bar{\Delta}(x)=-\frac{\pi a_s}{m}G_{\downarrow\downarrow,ii}(x,x).
\label{eq:LNld}
\end{align}
One notices that Eq.~(\ref{eq:LNlu}) and (\ref{eq:LNld}) coincide with the mean-field self-energies by the weak-coupling expansion shown in Eq.~(\ref{eq:S0uu}).

\section{Non-equilibrium dynamics}\label{sec:noeq}

The non-equilibrium dynamics of the spatially homogeneous dilute ultra-cold Fermi gas is investigated in real time on the Schwinger-Keldysh contour \cite{Schwinger:1960qe,Keldysh:1964ud}. The time evolution equations of two-point functions, called the Kadanoff-Baym equations, contain the integral of time that includes the memory of the past identified as the non-Markovian effect.

Starting from the identity of the Green's function
\begin{align}
\int_{z,\mathcal{C}}G_{\alpha\gamma,ik}^{-1}(x,z)G_{\gamma\beta,kj}(z,y)=\mathcal{I}_{\alpha\beta}\otimes\mathcal{I}_{ij}\,\delta_\mathcal{C}(x-y)\,,
\label{eq:Grid}
\end{align}
and plugging in Eq.~(\ref{eq:rrnG}), (\ref{eq:Gfre}) and (\ref{eq:Ginv}), the Kadanoff-Baym equations are obtained by extracting terms with or without ${\rm sgn}_{\mathcal{C}}$ given by
\begin{align}
&\,\bigg[ -\mathcal{I}_{\alpha\gamma}\otimes (\sigma_y)_{ik}i\partial_{x^0}+\mathcal{I}_{\alpha\gamma}\otimes \mathcal{I}_{ik}\frac{\nabla_\textbf{x}^2}{2m}-2M_{\alpha\gamma,ik}(x)\nonumber\\
&\,+i\Sigma_{\alpha\gamma,ik}(x,x)\bigg]F_{_{\gamma\beta,kj}}(x,y)\nonumber\\
=&\, \int d^{d-1}z\int_{t^0}^{y^0}\,dz^0\,\Sigma_{\alpha\gamma,ik}^F(x,z)\rho_{_{\gamma\beta,kj}}(z,y)\nonumber\\
&\,-\int d^{d-1}z\int_{t^0}^{x^0}\,dz^0 \,\Sigma_{\alpha\gamma,ik}^\rho(x,z)F_{_{\gamma\beta,kj}}(z,y),  \label{eq:EvFF}\\
&\, \bigg[ -\mathcal{I}_{\alpha\gamma}\otimes (\sigma_y)_{ik}i\partial_{x^0}+\mathcal{I}_{\alpha\gamma}\otimes \mathcal{I}_{ik}\frac{\nabla_\textbf{x}^2}{2m}-2M_{\alpha\gamma,ik}(x) \nonumber\\
&\,+i\Sigma_{\alpha\gamma,ik}(x,x)\bigg]
\rho_{_{\gamma\beta,kj}}(x,y)\nonumber\\
=&\,-\int d^{d-1} z \int_{y^0}^{x^0}\, dz^0\,\Sigma_{\alpha\gamma, ik}^\rho(x,z)\rho_{_{\gamma\beta,kj}}(z,y)\,,
\label{eq:EvRR}
\end{align}
where
\begin{align}
M_{\alpha\gamma,ik}(x)=
\begin{pmatrix}
\bar{\Delta}(x) & 0 \\
0 &  \Delta(x) 
\end{pmatrix}
_{\alpha\gamma}\otimes \mathcal{I}_{ik}
\label{eq:Mamm}
\end{align}
and any function labeled with an upper index $F$ or $\rho$ follows the decomposition in Eq.~(\ref{eq:rrnG}). In the $1/N$ expansion, decompositions of Eq.~(\ref{eq:SSuu}) and (\ref{eq:Piuu}) are given by
\begin{align}
\Sigma_{\alpha\beta,ij}^F(x,y)=&\,\frac{4\pi}{m}\bigg(F_{G_{\alpha^\prime\beta^\prime,ij}}(x,y)F_{D_{\alpha^\prime\beta^\prime}}(x,y)\nonumber\\
&\,-\frac{1}{4}\rho_{G_{\alpha^\prime\beta^\prime,ij}}(x,y)\rho_{D_{\alpha^\prime\beta^\prime}}(x,y)\bigg), \label{eq:sif1}\\
\Sigma_{\alpha\beta,ij}^\rho(x,y)=&\,\frac{4\pi}{m}\Big(F_{G_{\alpha^\prime\beta^\prime,ij}}(x,y)\rho_{D_{\alpha^\prime\beta^\prime}}(x,y)\nonumber\\
&\,+F_{D_{\alpha^\prime\beta^\prime}}(x,y)\rho_{G_{\alpha^\prime\beta^\prime,ij}}(x,y)\Big)\,,\label{eq:sif2}
\\
\Pi_{\alpha\beta}^F(x,y)=&\,\frac{2\pi}{m}\bigg(F_{G_{\alpha^\prime\beta^\prime,ij}}(x,y)F_{G_{\beta^\prime\alpha^\prime,ji}}(y,x)\nonumber\\
&\,+\frac{1}{4}\rho_{G_{\alpha^\prime\beta^\prime,ij}}(x,y)\rho_{G_{\beta^\prime\alpha^\prime,ji}}(y,x)\bigg)\,,\\
\Pi_{\alpha\beta}^\rho(x,y)=&\,\frac{2\pi}{m}\Big(\rho_{G_{\alpha^\prime\beta^\prime,ij}}(x,y)F_{G_{\beta^\prime\alpha^\prime,ji}}(y,x)\nonumber\\
&\,-F_{G_{\alpha^\prime\beta^\prime,ij}}(x,y)\rho_{G_{\beta^\prime\alpha^\prime,ji}}(y,x)\Big)\,,\label{eq:Piro}
\end{align}
where
\begin{align}
D_{\alpha\beta}(x,y)=F_{D_{\alpha\beta}}(x,y)-\frac{i}{2}\rho_{D_{\alpha\beta}}(x,y){\rm sgn}_\mathcal{C}(x^0-y^0)\,.
\end{align}

NLO corrections are scatterings beyond the mean-field level and expressed as a time integral that sums over all information during the evolution. I refer to Ref.~\cite{Berges:2004yj,berges2015nonequilibriumquantumfieldscold} for some detailed derivations in Eq.~(\ref{eq:EvFF}) and (\ref{eq:EvRR}) on the closed-time path.

Similarly, the same procedure can be done for $D_{\alpha\beta}(x,y)$ from Eq.~(\ref{eq:Grid}). So, relations analogous to Eq.~(\ref{eq:EvFF}) and (\ref{eq:EvRR}) are obtained as
\begin{align}
-\frac{i}{a_s}F_D(x,y)\sigma_x=&\,
i\int d^{d-1}z\int_{t^0}^{y^0}\,dz^0\,F_D(x,z)\Pi^\rho(z,y)\nonumber\\
&\,-i\int d^{d-1}z\int_{t^0}^{x^0}\,dz^0\,\rho_D(x,z)\Pi^F(z,y)\nonumber\\
&\,+i\frac{a_s}{N}\sigma_x\Pi^F(x,y)\,,\label{eq:FDFD} \\
-\frac{i}{a_s}\rho_D(x,y)\sigma_x 
=&\,
-i\int d^{d-1}z \int_{y^0}^{x^0}\,dz^0\,\rho_D(x,z)\Pi^\rho(z,y) \nonumber\\
&\,+i\frac{a_s}{N}\sigma_x\Pi^\rho(x,y)\,,
\label{eq:lDlD}
\end{align}
where the decomposition of the time-ordered two-point function in Eq.~(\ref{eq:rrnG}) is also applied to $D_{\alpha\beta}$ re-written by $F_{D_{\alpha\beta}}$ and $\rho_{D_{\alpha\beta}}$.

The retarded and advanced two-point function can be identified from the two-point function, i.e. spectral function, by
$D_R(x,y)\equiv\theta_\mathcal{C}(x^0-y^0)\rho_D(x,y)$ and $
D_A(x,y)\equiv-\theta_\mathcal{C}(y^0-x^0)\rho_D(x,y)$, respectively. They can be expressed in terms of central $X^\mu$ and relative coordinates $s^\mu$ given by $X_{xy}^\mu\equiv\big(x^\mu+y^\mu\big)/2$ and $s_{xy}^\mu\equiv x^\mu-y^\mu$, respectively. Consider the retarded and advanced part of Eq.~(\ref{eq:lDlD}) first by doing the Fourier transform with respect to $s_{xy}^\mu$ and performing the gradient expansion with $\partial_{X_{xy}^\mu}$ and $\partial_{p_\mu}$ given by
\begin{align}
&\,-\frac{i}{a_s}D_{R(A)}(X_{xy},p)\sigma_x-i\frac{a_s}{N}\sigma_x\Pi_{R(A)}(X_{xy},p)\nonumber\\
=&\,\int_{-\infty}^\infty ds^\nu e^{ip_\mu s_{xy}^\mu}\bigg[-\frac{i}{a_s}D_R\Big(X_{xy}+\frac{s_{xy}}{2},X-\frac{s_{xy}}{2}\Big)\sigma_x\nonumber\\
&\,-i\frac{a_s}{N}\sigma_x\Pi_R\Big(X_{xy}+\frac{s_{xy}}{2},X_{xy}-\frac{s_{xy}}{2}\Big)\bigg]\nonumber\\
=&\,-i\int_{-\infty}^\infty d^dz\int_{-\infty}^\infty ds_{xy}\int_{-\infty}^\infty dk \int_{-\infty}^\infty dq \nonumber\\
&\,\times e^{ip_\mu s_{xy}^\mu-k_\mu s_{sz}^\mu -q_\mu s_{zy}^\mu}D_{R(A)}(X_{xz},k)\Pi_{R(A)}(X_{zy},q)\,,
\label{eq:lolo}
\end{align}
where $\Pi_{R(A)}$ is defined by $\Pi^\rho$ similar to $D_{R(A)}$. The lead-order terms in
\begin{align}
D_{R(A)}(X_{xz},k)=&\, {\rm exp}\bigg(\frac{s_{zy}^\mu}{2}\frac{\partial}{\partial X_{xy}^\mu}\bigg)D_R(X_{xy},k)\,,\\
\Pi_{R(A)}(X_{zy},q)=&\,{\rm exp}\bigg(-\frac{s_{xz}^\mu}{2}\frac{\partial}{\partial X_{xy}^\mu}\bigg)\Pi_R(X_{xy},q)
\end{align}
are kept in Eq.~(\ref{eq:lolo}) such that Eq.~(\ref{eq:lolo}) reduces to
\begin{align}
&\,-\frac{N}{a_s}D_{R(A)}(X_{xy},p)\Big(\sigma_{x}-a_s \Pi_{R(A)}(X_{xy},p)\Big)\nonumber\\
=&\,a_s\sigma_{x}\Pi_{R(A)}(X_{xy},p)
\label{eq:rdfd}
\end{align}
in the leading-order gradient expansion. A similar procedure can be applied in Eq.~(\ref{eq:EvFF}), (\ref{eq:EvRR}) and (\ref{eq:FDFD}) under the limit $t^0\rightarrow-\infty$. It is found that Eq.~(\ref{eq:FDFD}) is obtained with a similar form of Eq.~(\ref{eq:rdfd}) as
\begin{align}
&\,-\frac{N}{a_s}D_{R(A)}^F(X_{xy},p)\Big(\sigma_{x}-a_s \Pi_{R(A)}^F(X_{xy},p)\Big)\nonumber\\
=&\,a_s\sigma_{x}\Pi_{R(A)}^F(X_{xy},p)\,,
\label{eq:ddfd}
\end{align}
where $D_{R(A)}^F$ and $\Pi_{R(A)}^F$ are retarded (advanced) $F_D$ and $\Pi^F$, respectively. Since only $X_{xy}^\mu$-dependence shows up in the leading-order gradient expansion, I will express $X_{xy}^\mu$ as $X^\mu$ in the following.

Two-point functions $\rho_{D}(X,p)$ and $F_D(X,p)$ can be obtained as
\begin{align}
-\frac{N}{a_s}\rho_{D}(X,p)=&\,a_s\bigg(\sigma_{x}+a_s\sigma_{x}\Pi_{R}(X,p)K_{R}(X,p)\bigg)\nonumber\\
&\,\times\Pi^\rho(X,p)K_{A}(X,p)\,,
\label{eq:rDrD}\\
-\frac{N}{a_s}F_{D}(X,p)=&\,a_s\Big(\sigma_{x}+a_s\sigma_{x}\Pi_R(X,p)K_{R}(X,p)\Big)\nonumber\\
&\,\times\Pi^F(X,p)K_{A}(X,p)\,,
\label{eq:DfDf}
\\
K_{R(A)}^{-1}(X,p)\equiv&\,\sigma_x-a_s\Pi_{R(A)}(X,p)\,,
\end{align}
by the relation $D_{R}(X,p)-D_A(X,p)=\rho_D(X,p)$ and $D_R^F(X,p)-D_A^F(X,p)=F_D(X,p)$. Their expressions with a compact notation are given by
\begin{align}
&\,\rho_{D_{\alpha^\prime\alpha^\prime}}(X,p)\nonumber\\=&-\frac{a_s^2}{N}\Big(v(X,p)\Pi_{\bar{\alpha}^\prime\bar{\alpha}^\prime}^\rho(X,p)+\bar{v}_{\alpha^\prime\alpha^\prime}(X,p)\Pi_{\alpha^\prime\alpha^\prime}^\rho(X,p)\Big), \label{eq:rDuu}\\
&\,F_{D_{\alpha^\prime\alpha^\prime}}(X,p)\nonumber\\=&-\frac{a_s^2}{N}\Big(v(X,p)\Pi_{\bar{\alpha}^\prime\bar{\alpha}^\prime}^F(X,p)+\bar{v}_{\alpha^\prime\alpha^\prime}(X,p)\Pi_{\alpha^\prime\alpha^\prime}^F(X,p)\Big),\label{eq:FDdd}
\end{align}
with
\begin{align}
v(X,p)\equiv&\,\big(a_s^2\Pi_{A_{\uparrow\uparrow}}(X,p)\Pi_{A_{\downarrow\downarrow}}(X,p)-1\big)^{-1}\nonumber\\
&\,\times\big(a_s^2\Pi_{R_{\uparrow\uparrow}}(X,p)\Pi_{R_{\downarrow\downarrow}}(X,p)-1\big)^{-1},\label{eq:vcor}\\
\bar{v}_{\alpha^\prime\alpha^\prime}(X,p)\equiv&\,a_s^2\Pi_{R_{\bar{\alpha}^\prime\bar{\alpha}^\prime}}(X,p)\Pi_{A_{\bar{\alpha}^\prime\bar{\alpha}^\prime}}(X,p)\,v(X,p), \label{eq:vbou}
\end{align}
by Eq.~(\ref{eq:rDrD}) and (\ref{eq:DfDf}). Note that I only show diagonal components that are non-trivial at the NLO in the $1/N$ expansion. One can see that Eq.~(\ref{eq:vcor}) and (\ref{eq:vbou}) play a role of corrections to the coupling in Eq.~(\ref{eq:rDuu}) and (\ref{eq:FDdd}) by resumming infinite numbers of the fermion loop at the four-vertex. That is, the coupling receives non-local contributions that do not take place in the weak-coupling expansion.

The gradient expansion of Eq.~(\ref{eq:EvFF}) and (\ref{eq:EvRR}) can help derive the transport equations describing the transport of conserved quantities at the late time of the evolution by taking the limit $t^0\rightarrow-\infty$ \cite{kadanoff1962quantum,PhysRevD.37.2878,Berges_2006}. To obtain transport equations, the time evolution of $F_{\alpha\beta,ij}(x,y)$ formulated by Eq.~(\ref{eq:EvFF}) is considered first to minus itself with $x\leftrightarrow y$ (\ref{eq:EvRR}) by taking the first order in $\partial_{X^\mu}$ after performing the Fourier transform with respect to relative coordinates $s^\mu$. The procedure employs the cyclic property of trace and symmetry properties of two-point functions shown in Eq.~(\ref{eq:FRRR}) and (\ref{eq:rrgg}). Similarly, the above process is applied to Eq.~(\ref{eq:EvRR}) by only replacing the subtraction with the addition. The resulting transport equations are obtained as
\begin{align}
\frac{\partial}{\partial X^0}{\rm Tr}\,\Big[\mathcal{I}\otimes\sigma_y\,F(X,p)\Big] =&\,{\rm Tr}\Big[\tilde{\Sigma}^\rho(X,p)F(X,p)\nonumber\\
&\,-\Sigma^F(X,p)\tilde{\rho}(X,p)\Big]\,, 
\label{eq:Ftre} \\
\frac{\partial }{\partial X^0}{\rm Tr}\,\Big[\mathcal{I}\otimes\sigma_y\, \tilde{\rho}(X,p) \Big]=&\,0\,,
\label{eq:trrr}
\end{align}
with $\tilde{\rho}(X,p)\equiv -i\rho(X,p)$ and $\tilde{\Sigma}^\rho(X,p)\equiv-i\Sigma^\rho(X,p)$.  The spatial homogeneity of the system implies that the two-point function is independent of $\textbf{X}$, so only $\partial_{X^0}$ remains. It is found that the late-behavior of the spectral function becomes independent of $X^0$ at least at the first order of gradient expansion.

The Kadanoff-Baym ansatz for the fermion is given by \cite{Berges_2003,Branschadel:2008sk,Gasenzer_2009}
\begin{align}
F_{_{\alpha\beta,ij}}(X,p)=\bigg(\frac{1}{2}-f(X,p)\bigg)\tilde{\rho}_{_{\alpha\beta,ij}}(X,p),
\label{eq:Fsyr}
\end{align}
which reduces to the fluctuation-dissipation theorem if $f(X,p)$ is the Fermi-Dirac distribution. The on-shell particle number can be obtained by fixing a dispersion relation for $f(X,p)$ with the help of the quasi-particle assumption, where the spectral function in the free-field form with a vanishing width \cite{Branschadel:2008sk}
\begin{align}
&\,\tilde{\rho}_{_{\alpha\beta,ij}}(X,p^0,\textbf{p})\nonumber\\
=&\,-\pi\bigg[\delta\bigg(p^0+\frac{\textbf{p}^2}{2m}\bigg)-\delta\bigg(p^0-\frac{\textbf{p}^2}{2m}\bigg)\bigg]\mathcal{I}_{\alpha\beta}\otimes\mathcal{I}_{ij}\nonumber\\
&\,-\pi\bigg[\delta\bigg(p^0+\frac{\textbf{p}^2}{2m}\bigg)+\delta\bigg(p^0-\frac{\textbf{p}^2}{2m}\bigg)\bigg]\mathcal{I}_{\alpha\beta}\otimes(\sigma_y)_{ij}\,, \label{eq:frsp}
\end{align}
is imposed. 
Plugging Eq.~(\ref{eq:Fsyr}) into Eq.~(\ref{eq:Ftre}) with the help of Eq.~(\ref{eq:rrnG}) and (\ref{eq:SSuu})(\ref{eq:Piuu})(\ref{eq:rDuu})(\ref{eq:FDdd}) in the momentum space, the transport equation in the $1/N$ expansion is given by
\begin{widetext}
\begin{align}\frac{\partial f(t,\textbf{p})}{\partial t}=&\,-\frac{8\pi^2 a_s^2}{Nm^2}\int_0^\infty\frac{dp^0}{2\pi}\int_{q,l,r}(2\pi)^d\delta^d(p+l-q-r)\nonumber\\
&\,\times\Big(v(t,p-q)\tilde{\rho}_{\bar{\alpha}\bar{\alpha},ks}(t,r)\tilde{\rho}_{\bar{\alpha}\bar{\alpha},sk}(t,l)+\bar{v}_{\alpha\alpha}(t,p-q)\tilde{\rho}_{\alpha\alpha,ks}(t,r)\tilde{\rho}_{\alpha\alpha,sk}(t,l)\Big)\tilde{\rho}_{\alpha\alpha,ij}(t,q)\tilde{\rho}_{\alpha\alpha,ji}(t,p)\nonumber\\
&\,\times\bigg(\Big(1-f(t,r)\Big)f(t,l)\Big(1-f(t,q)\Big)f(t,p)-f(t,r)\Big(1-f(t,l)\Big)f(t,q)\Big(1-f(t,p)\Big)\bigg),
\label{eq:trfe}
\end{align}
\end{widetext}
where $X^0\equiv t$ and $\int_q\equiv\int_{-\infty}^{\infty}d^dq/(2\pi)^d$ ($\int_\textbf{q}\equiv\int_{-\infty}^\infty d^{d-1}\textbf{q}/(2\pi)^{d-1}$); the effective particle number distribution is defined
\begin{align}
f(t,\textbf{p})=-\int_0^\infty \frac{dp^0}{2\pi}{\rm Tr}\big[\mathcal{I}\otimes\sigma_y\tilde{\rho}(t,p)\big]f(t,p).
\label{eq:efnu}
\end{align}
Note that Eq.~(\ref{eq:trfe}) goes beyond the kinetic equation since frequency integrals are still present and the spectral function is not necessary to be assumed as the free-field form in Eq.~(\ref{eq:frsp}), which fixes a dispersion relation \cite{Scheppach:2009wu,berges2015nonequilibriumquantumfieldscold}.

On the other hand, the transport equation by the weak-coupling expansion is easily obtained by the limit $v(X,p-q)\rightarrow 1$ and $\bar{v}_{\alpha^\prime\alpha^\prime}(X,p-q)\rightarrow0$ as $|a_s\Pi_{R(A)_{\alpha\beta}}|\ll1$.

\section{Stationary solutions with scale-invariant flux}
Some stationary solutions are characterized as non-thermal states related to scale-invariant fluxes determined by scaling exponents \cite{Zakharov:2013ko}. To begin with, I first revisit the characteristics of non-thermal states and the corresponding scale-invariant flux related to the weak-coupling case in the following. Then, the case with a large $a_s$ will be discussed by doing the $1/N$ expansion. It will be seen in Sec.~\ref{sec:Kolm} that these scaling structures of the effective particle number distribution related to scale-invariant flux correspond to stationary solutions of the kinetic equation.

The continuity equations for the on-shell particle number and the energy are given by
\begin{align}
\frac{\partial }{\partial t}\big(\omega(\textbf{p})f(t,\textbf{p})\big)+\nabla \cdot \tilde{\textbf{j}}(\textbf{p})=&\,0.
\label{eq:Conc}
\end{align}
The dispersion relation is given by $\omega(\textbf{p})=|\textbf{p}|^\tau$, where $\tau=0$ $(\tau=z)$ corresponds to the particle (energy) in Eq.~(\ref{eq:Conc}). The integral of the divergence, according to the Gauss' law, is identical to the surface integral, which gives the flux $\tilde{A}(k)$
\begin{align}
\int_{\textbf{p}}^k\nabla\cdot \tilde{\textbf{j}}(\textbf{p})=\int_{\partial k}\tilde{\textbf{j}}(\textbf{p})\cdot d\textbf{A}\equiv&\,(2\pi)^{d-1}\tilde{A}(k),
\end{align}
where
\begin{align}
\tilde{A}(k)=&\,\Omega\int^k d|\textbf{p}| |\textbf{p}|^{d-2}\omega(\textbf{p})\mathcal{C}(\textbf{p})\,,\\
\Omega\equiv&\, -\frac{1}{2^{d-1}\pi^{\frac{d-1}{2}}\Gamma\big(\frac{d-1}{2}+1\big)}\,,
\end{align}
and $\mathcal{C}(\textbf{p})$ is the collision kernel given by
\begin{align}
\frac{\partial f(t,\textbf{p})}{\partial t}=\mathcal{C}(\textbf{p}).
\end{align}
The scaling of the distribution function denotes $f(t,\textbf{p})=s^{\kappa_w}f(t,s\textbf{p})$ 
and $\mathcal{C}(\textbf{p})$ scales as
\begin{align}
\mathcal{C}(\textbf{p})=&\,s^{-\mu}\mathcal{C}(s\textbf{p})\,,
\label{eq:caca}\\
-\mu=&\,-3(d-1)+z+d-1+n\kappa_w \nonumber\\
=&\,-2d+2+z+n\kappa_w\,,
\end{align}
where Eq.~(\ref{eq:trfe}) is taken to be on-shell and $n$ denotes the number of $f$ in a product contained in the collision kernel. By setting $s=|\textbf{p}|^{-1}$, the momentum-dependence of the collision kernel can be totally scaled out of itself
\begin{align}
\mathcal{C}(\textbf{p})=|\textbf{p}|^{2d-2-z-n\kappa_w}\mathcal{C}(1).
\end{align}
Therefore, the flux of the particle and the energy can be expressed as
\begin{align}
\tilde{A}(k)=&\,\Omega\int^k d|\textbf{p}| |\textbf{p}|^{\mu+d-2+\tau}\omega(1)\mathcal{C}(1)
\label{eq:tiAk}\\
\sim&\,\frac{k^{2d-2-z-n\kappa_w+d-2+\tau+1}}{2d-2-z-n\kappa_w+d-2+\tau+1}\mathcal{C}(1),
\label{eq:flua}
\end{align}
which is scale-invariant if the exponent vanishes
\begin{align}
3d-3-z-n\kappa_w=&\,0\,, \quad \tau=0\,,\\
3d-3-n\kappa_w=&\,0\,, \quad \tau=z\,.
\end{align}
Hence, scaling exponents have solutions given by
\begin{align}
\kappa_w=&\,\frac{3d-z-3}{n}\,, \quad \tau=0\,, \label{eq:kwpa}\\
\kappa_w=&\,\frac{3d-3}{n}\,,\quad \quad\,\,\,\,\tau=z\,,
\label{eq:kwen}
\end{align}
and they reproduce results in Ref.~\cite{PismaZhETF.21.13,Kats:1976,Zakharov:2013ko} when $n=2$ related to the Boltzmann equation.

Note that the above derivations from Eq.~(\ref{eq:Conc}) to Eq.~(\ref{eq:tiAk}) can be generalized to the case expressed in terms of the effective particle number in Eq.~(\ref{eq:efnu}), which re-writes the energy distribution $\omega(\textbf{p})f(t,\textbf{p})$ as
\begin{align}
-\int_0^\infty \frac{dp^0}{2\pi}{\rm Tr}\big[\mathcal{I}\otimes\sigma_y\tilde{\rho}(t,p)\big]p^0f(t,p).
\label{eq:efen}
\end{align}

For general cases, the scaling form of the spectral function is considered
\begin{align}
\rho_{\alpha\beta,ij}(t,p^0,\textbf{p})=s^{2-\eta}\rho_{\alpha\beta,ij}(t,s^zp^0,s\textbf{p})\,,
\label{eq:scro}
\end{align}
where $z$ is the dynamical scaling exponent and $\eta$ is the anomalous dimension assumed to be tiny in this study. The function $f(t,p^0,\textbf{p})$ scales as $f(t,p^0,\textbf{p})=s^{\kappa_s}f(t,s^zp^0,s\textbf{p})$ and the effective particle number distribution defined in Eq.~(\ref{eq:efnu})
has a scaling form $f(t,\textbf{p})\sim |\textbf{p}|^{-\kappa_s+z-2+\eta}$. In the case with large $a_s$, the vertex correction in Eq.~(\ref{eq:vcor}) derived from $1/N$ expansion becomes significant, which contributes an extra scaling 
\begin{align}
\Pi_{R(A)_{\alpha\beta}}(t,p^0,\textbf{p})=s^{\kappa_s-z-(d-1)+2(2-\eta)}\Pi_{R(A)_{\alpha\beta}}(t,s^zp^0,s\textbf{p})\label{eq:Pisc}
\end{align}
by the scaling in the retarded (advanced) part of Eq.~(\ref{eq:Piro}) from Eq.~(\ref{eq:Fsyr}) and (\ref{eq:scro}).

Based on Eq.~(\ref{eq:vcor}) and (\ref{eq:vbou}), the scaling behavior will be dominated by $\bar{v}_{\alpha^\prime\alpha^\prime}(t,p)$ over $v(t,p)$ if $|a_s\Pi_{R(A)_{\alpha\beta}}(t,p)|\gg1$. The significant correction results in the scaling exponent $-\mu$ defined in Eq.~(\ref{eq:caca}) of the kinetic equation Eq.~(\ref{eq:trfe}) as
\begin{align}
-\mu=&\,-4z-3(d-1)+z+d-1+4(2-\eta)\nonumber\\
&\,-2\big(\kappa_s-z-(d-1)+2(2-\eta)\big)+n\kappa_s \nonumber\\
=&\,-z+(n-2)\kappa_s\,,
\label{eq:muco}
\end{align}
where $n$ is the number of $f$ in a product in the collision kernel. One obtains the relations
\begin{align}
d-2+z-(n-2)\kappa_s+1=&\,0\,,\quad \tau=0,\label{eq:rec1}\\
d-2+z-(n-2)\kappa_s+z+1=&\,0\,,\quad \tau=z,
\label{eq:rec2}
\end{align}
by Eq.~(\ref{eq:tiAk}) for the scale-invariant flux.

\section{Kolmogorov-type non-thermal fixed points and beyond}\label{sec:Kolm}

Stationary solutions are solutions of two-point functions $F_{_{\alpha\beta,ij}}(x,y)$ and $\rho_{\alpha\beta,ij}(x,y)$ in Eq.~(\ref{eq:EvFF}) and (\ref{eq:EvRR}) that have translational invariance in space and time obtained by solving the kinetic equation, where the $X^0$-dependence is lost and transport equations in Eq.~(\ref{eq:Ftre}) and (\ref{eq:trrr}) become vanishing. The stationary condition for this kind of solution can be derived, with the help of Eq.~(\ref{eq:FRRR}) and (\ref{eq:rrgg}), by considering the trace of LHS of Eq.~(\ref{eq:EvFF}) and (\ref{eq:EvRR}). The trace of RHS and itself under the exchange $x\leftrightarrow y$, on the other hand, are subtracted for the evolution of $F(X,p)$ or added for $\rho(X,p)$, in which the upper limit of the integrals corresponding to $F(X,p)$ extends to infinity while the integral vanishes in the case of $\rho(X,p)$. Therefore, the stationary condition is satisfied by the vanishing trace of the memory integrals in Kadanoff-Baym equations since the trace of LHS of Eq.~(\ref{eq:EvFF}) is invariant while that of Eq.~(\ref{eq:EvRR}) is up to a minus sign after $x\leftrightarrow y$. The condition in the momentum space leads to
\begin{align}
{\rm Tr}\Big[\tilde{\Sigma}^\rho(X,p)F(X,p)-\Sigma^F(X,p)\tilde{\rho}(X,p)\Big]=0
\label{eq:stcc}
\end{align}
in the transport equation at the late time for the existence of stationary solutions, while the spectral function satisfies it trivially by Eq.~(\ref{eq:trrr}). The stationary condition Eq.~(\ref{eq:stcc}) is consistent with the vanishing transport equations in Eq.~(\ref{eq:Ftre}) and (\ref{eq:trrr}) due to the absence of $X^0$-dependence in the stationary solutions.

In this section, a slightly different stationary condition is considered by integrating over spatial momenta $\textbf{p}$ of Eq.~(\ref{eq:stcc})
\begin{align}
\int_\textbf{p}{\rm Tr}\Big[\tilde{\Sigma}^\rho(X,p)F(X,p)-\Sigma^F(X,p)\tilde{\rho}(X,p)\Big]=0\,.
\label{eq:stco}
\end{align}
In the physical sense, all collisions in total turn out be collisionless after averaging over $\textbf{p}$. Therefore, the scaling transformation \cite{Scheppach:2009wu}, shown in Appendix~\ref{sec:sctr}, is feasible by a suitable change of variables to exchange the external frequency $p^0$ with other frequencies as integration variables in the kinetic equation, which is analogous to the case of weak wave turbulence where the Zakharov transformation is applied \cite{Zakharov:1992za}.

This kind of stationary solutions are found in this section to compare with stationary solutions in Eq.~(\ref{eq:kwpa})(\ref{eq:kwen}). For simplicity in the notation, any momentum dependence is moved to the lower index and the notion of time variable is suppressed due to Eq.~(\ref{eq:trrr}). The indices with $\uparrow\uparrow$ and $\downarrow\downarrow$ are also ignored for the spin-balanced case considered in this study.

By the stationary condition in Eq.~(\ref{eq:stco}), it is required to check whether stationary solutions satisfy
\begin{align}
&-\frac{8}{N}\bigg(\frac{\pi a_s}{m}\bigg)^2\int_\textbf{p}\int_{q,l,r}(2\pi)^d\delta_{p+l-q-r}\,\tilde{\rho}_{mn,r}\tilde{\rho}_{nm,l}\tilde{\rho}_{ij,q}\tilde{\rho}_{ji,p}\nonumber\\
&\Big(\big(1-f_r\big)f_l\big(1-f_q\big)f_p-f_r\big(1-f_l\big)f_q\big(1-f_p\big)\Big)=0\,,
\label{eq:poco}
\end{align}
where $\delta_{p+l-q-r}\equiv\delta^d(p+l-q-r)$, in the perturbation theory reduced from Eq.~(\ref{eq:trfe}). Next, by the change of variables, all the negative sides of the frequency integrals are transformed into their positive intervals by the rotational invariance and $f_{-p}=1-f_p$ given by Eq.~(\ref{eq:FRRR}), (\ref{eq:rrgg}) and (\ref{eq:Fsyr}). The transport equation turns out to be the form of kinetic equation. The consequent integral enables one to use the scaling transformation, shown in Appendix~\ref{sec:sctr}, employed to interchange momenta $p$ with all the other momenta $q$, $l$, $r$ in the next step. Hence, the integral in Eq.~(\ref{eq:poco}) finally turns out to consist of terms proportional to $f$ denoted by $\mathcal{O}(f^1)$ and $f^2$ by $\mathcal{O}(f^2)$

\begin{widetext}
\begin{align}
\mathcal{O}(f^1):
&\,-\frac{2}{N}\bigg(\frac{\pi a_s}{m}\bigg)^2\int_0^\infty dq^0\,dl^0\, d r^0\int_{\textbf{p},\textbf{l},\textbf{q},\textbf{r}}(p^0)^{-\bar{\Delta}}\bigg[\nonumber\\
&\,-\tilde{\rho}_{mn,r}\,\tilde{\rho}_{nm,l}\,\tilde{\rho}_{ij,p}\,\tilde{\rho}_{(3-i)(3-j),q}(-1)^{i+j+1}f_r\delta_{p+l+q-r}\Big((p^0)^{\bar{\Delta}}+(l^0)^{\bar{\Delta}}+(q^0)^{\bar{\Delta}}-(r^0)^{\bar{\Delta}}\Big)\nonumber\\
&\,-\tilde{\rho}_{mn,p}\,\tilde{\rho}_{nm,q}\,\tilde{\rho}_{ij,r}\,\tilde{\rho}_{(3-i)(3-j),l}(-1)^{i+j+1}f_q\delta_{p+l-q+r}\Big((p^0)^{\bar{\Delta}}+(l^0)^{\bar{\Delta}}-(q^0)^{\bar{\Delta}}+(r^0)^{\bar{\Delta}}\Big)\nonumber\\
&\,-\tilde{\rho}_{mn,r}\,\tilde{\rho}_{nm,l}\,\tilde{\rho}_{ij,p}\,\tilde{\rho}_{(3-i)(3-j),q}(-1)^{i+j+1}f_l\delta_{p-l+q+r}\Big((p^0)^{\bar{\Delta}}-(l^0)^{\bar{\Delta}}+(q^0)^{\bar{\Delta}}+(r^0)^{\bar{\Delta}}\Big) \nonumber\\
&\,+\tilde{\rho}_{mn,p}\,\tilde{\rho}_{nm,q}\,\tilde{\rho}_{ij,r}\,\tilde{\rho}_{(3-i)(3-j),l}(-1)^{i+j+1}f_p\delta_{p-l-q-r}\Big((p^0)^{\bar{\Delta}}-(l^0)^{\bar{\Delta}}-(q^0)^{\bar{\Delta}}-(r^0)^{\bar{\Delta}}\Big)\bigg]\,,\label{eq:JJon}\\
\nonumber\\
\mathcal{O}(f^2):
&\,-\frac{2}{N}\bigg(\frac{\pi a_s}{m}\bigg)^2\int_0^\infty dq^0\,dl^0\, d r^0\int_{\textbf{p},\textbf{l},\textbf{q},\textbf{r}}(p^0)^{-\Delta}\bigg[\nonumber\\
&\,-\tilde{\rho}_{mn,p}\,\tilde{\rho}_{nm,q}\,\tilde{\rho}_{ij,r}\,\tilde{\rho}_{(3-i)(3-j),l}(-1)^{i+j+1}f_p\big(f_l+f_q+f_r\big)\delta_{p-l-q-r}\Big((p^0)^\Delta-(l^0)^\Delta-(q^0)^\Delta-(r^0)^\Delta\Big)\nonumber\\
&\,+\tilde{\rho}_{mn,r}\,\tilde{\rho}_{nm,l}\,\tilde{\rho}_{ij,p}\,\tilde{\rho}_{(3-i)(3-j),q}(-1)^{i+j+1}f_r\big(f_p+f_l+f_q\big)\delta_{p+l+q-r}\Big((p^0)^\Delta+(l^0)^\Delta+(q^0)^\Delta-(r^0)^\Delta\Big)\nonumber\\
&\,+\tilde{\rho}_{mn,p}\,\tilde{\rho}_{nm,q}\,\tilde{\rho}_{ij,r}\,\tilde{\rho}_{(3-i)(3-j),l}(-1)^{i+j+1}f_q\big(f_p+f_l+f_r\big)\delta_{p+l-q+r}\Big((p^0)^\Delta+(l^0)^\Delta-(q^0)^\Delta+(r^0)^\Delta\Big)\nonumber\\
&\,+\tilde{\rho}_{mn,p}\,\tilde{\rho}_{(3-m)(3-n),q}\,\tilde{\rho}_{ij,r}\,\tilde{\rho}_{(3-i)(3-j),l}(-1)^{m+n+i+j}\big(f_pf_q-f_lf_r\big)\delta_{p-l+q-r}\Big((p^0)^\Delta-(l^0)^\Delta+(q^0)^\Delta-(r^0)^\Delta\Big)\nonumber\\
&\,+\tilde{\rho}_{mn,r}\,\tilde{\rho}_{nm,l}\,\tilde{\rho}_{ij,p}\,\tilde{\rho}_{(3-i)(3-j),q}(-1)^{i+j+1}f_l\big(f_p+f_q+f_r\big)\delta_{p-l+q+r}\Big((p^0)^\Delta-(l^0)^\Delta+(q^0)^\Delta+(r^0)^\Delta\Big)\nonumber\\
&\,+\tilde{\rho}_{mn,p}\,\tilde{\rho}_{nm,q}\,\tilde{\rho}_{ij,r}\,\tilde{\rho}_{ji,l}\big(f_pf_r-f_lf_q\big)\delta_{p-l-q+r}\Big((p^0)^\Delta-(l^0)^\Delta-(q^0)^\Delta+(r^0)^\Delta\Big)\nonumber\\
&\,+\tilde{\rho}_{mn,p}\,\tilde{\rho}_{nm,q}\,\tilde{\rho}_{ij,r}\,\tilde{\rho}_{ji,l}\big(f_pf_l-f_qf_r\big)\delta_{p+l-q-r}\Big((p^0)^\Delta+(l^0)^\Delta-(q^0)^\Delta-(r^0)^\Delta\Big)\bigg]\,,
\label{eq:JJse}
\end{align}
\end{widetext}
where 
\begin{align}
\bar{\Delta}=&\,-\frac{1}{z}\Big(3(d-1)+3z-\kappa_w-4(2-\eta)\Big)\,,\\
\Delta=&\,-\frac{1}{z}\Big(3(d-1)+3z-2\kappa_w-4(2-\eta)\Big)\,,
\end{align}
to the quadratic order of $f$ in the collision kernel.

One can easily find possible stationary solutions $\bar{\Delta}=1\,,\,\Delta=1$ that satisfy vanishing Eq.~(\ref{eq:JJon}) and (\ref{eq:JJse}) based on the constraints imposed by Dirac delta functions. However, this set of solutions does not give a self-consistent result since
\begin{align}
&\,\bar{\Delta}=1\quad\Rightarrow\quad \kappa_w=-11+3d+4z+4\eta\,,\label{eq:bdel}\\
&\,\Delta=1\quad\Rightarrow\quad \kappa_w=\frac{1}{2}\big(-11+3d+4z+4\eta\big).
\label{eq:enff}
\end{align}
The resolution is the quasi-particle assumption by inserting the spectral function in the free-field form in Eq.~(\ref{eq:frsp}), which implies that $\tilde{\rho}_{ij,p}\,\tilde{\rho}_{(3-i)(3-j),q}(-1)^{i+j+1}$ is vanishing with a fixed dispersion relation. Therefore, the constraint of $\bar{\Delta}$ is lifted and only the value of $\Delta$ matters for stationary solutions. Remarkably, this argument is consistent with the particle nature of the Boltzmann equation. In this sense, the solution given by Eq.~(\ref{eq:enff}) is allowed and corresponds to the scale-invariant energy flux given in Eq.~(\ref{eq:kwen}) if $n=2$, $z=2$ and $\eta=0$.

One can notice that multiple terms in Eq.~(\ref{eq:JJon}) and (\ref{eq:JJse}) are removed by the quasi-particle assumption except the last two terms in Eq.~(\ref{eq:JJse}) because $\tilde{\rho}_{mn,p}\,\tilde{\rho}_{nm,q}$ is non-vanishing even by Eq.~(\ref{eq:frsp}). They can be eliminated by
\begin{align}
\Delta=0\,
\quad\Rightarrow\quad\kappa_w=\frac{1}{2}\big(-11+3d+3z+4\eta\big)\,,
\label{eq:Dezz}
\end{align}
which reproduces the result related to the scale-invariant particle flux given by Eq.~(\ref{eq:kwpa}) if $n=2$, $z=2$ and $\eta=0$. In summary, Eq.~(\ref{eq:enff}) and (\ref{eq:Dezz}) reproduce Kolmogorov non-thermal fixed points for the Boltzmann equation when the dilute ultra-cold Fermi gas system is on-shell and described by the perturbation theory. In addition, they show the nature different from the ultra-cold Bose gas \cite{Scheppach:2009wu}, where the scale-invariant particle flux is fulfilled by the quasi-particle assumption while the scale-invariant energy flux avoids this restriction. The difference demonstrates the particle and wave interactions in two systems, respectively.

The removal of linear terms in $f$ by the quasi-particle assumption is related to the standard Boltzmann equation recovered in the perturbation theory in the on-shell limit when $f\ll1$, where Eq.~(\ref{eq:JJon}) is kinematically disfavored. On the other hand, it would be tempting to consider Eq.~(\ref{eq:bdel}) a valid perturbative stationary solution corresponding to the scale-invariant energy flux beyond the Kolmogorov-type non-thermal fixed point without using the form in Eq.~(\ref{eq:frsp}), in which the stationary condition is only considered to the order in Eq.~(\ref{eq:JJon}) when $f\ll1$. If $\mathcal{O}(f)$ dominates over $\mathcal{O}(f^2)$, the energy flux by Eq.~(\ref{eq:bdel}) is fixed only by the energy conservation, analogous to the case in \cite{Scheppach:2009wu}.

The vertex corrections to Eq.~(\ref{eq:JJse}) lead to a solution
\begin{align}
\Delta=&\,-\frac{1}{z}\Big[3(d-1)+3z-n\kappa_s-4(2-\eta)\nonumber\\
&\,+2\Big(\kappa_s-z-(d-1)+2(2-\eta)\Big)\Big]\,,
\end{align}
which justifies relations in Eq.~(\ref{eq:rec1}) and (\ref{eq:rec2}). However, scaling solutions beyond Kolmogorov-type non-thermal fixed points by the $1/N$ expansion are found not possible when the quasi-particle assumption is used, where the leading terms in the collision kernel fix $n=2$. Resummed diagrams with $|a_s\Pi_{{R(A)}_{\alpha\beta}}(t,p)|\gg1$ result in a cancelation of $\kappa_s$ in the scaling structure of the collision kernel in Eq.~(\ref{eq:muco}). Two conditions for scale-invariant flux in Eq.~(\ref{eq:rec1}) and (\ref{eq:rec2}) are spoiled since both $d-1$ and $z$ are positive.

\section{Conclusions}
\label{sec:conc}
The real-time dynamics of non-equilibrium dilute ultra-cold Fermi gas and its Kolmogorov-type non-thermal fixed points and beyond are investigated by including the scattering process from the NLO quantum corrections. Two types of expansion schemes are performed for the case of weak and moderate coupling.

The Kolmogorov-type non-thermal fixed points are characterized by the stationary condition, where the scaling form of the distribution function and the spectral function by momenta are assumed. The technique of scaling transformation allows one to find the specific scaling exponents that fulfill the stationary condition. The quasi-particle assumption applied in the kinetic equation is found to reproduce the scaling exponents derived in earlier studies by Kats, Kontorovich, Moiseev and Novikov \cite{PismaZhETF.21.13,Kats:1976}. In essence, the Boltzmann equation accounts for interactions between particles. That is why both conservation laws rely on the quasi-particle assumption in the perturbation theory. They are distinct from the weak wave turbulence of the ultra-cold Bose gas, where only the energy conservation can be independent of the quasi-particle assumption \cite{Scheppach:2009wu}.

The kinetic equation beyond the Boltzmann equation by $1/N$ expansion enables one to look for stationary solutions beyond the Kolmogorov-type non-thermal fixed points. After adopting the quasi-particle assumption, the scaling $\kappa_s$ of particle's distribution in the collision kernel is found to cancel with each other due to the additional scaling originating from non-negligible resummed diagrams that the scale-invariant flux of particle and energy do not exist. It is distinct from the strong wave turbulence induced also by the resummed diagrams in the IR regime of the ultra-cold Bose gas \cite{Berges_2008}.

\section{Acknowledgments}
I would like to thank Paul Romatschke for discussions and Johannes Reinking for valuable comments. This work is partially supported by TASI Summer 2025.

\appendix

\section{Scaling transformation}
\label{sec:sctr}
Following calculations in Appendix.~F in \cite{Scheppach:2009wu}, the integral has an identity
\begin{align}
&\,\int_\textbf{p}\int_{l,q,r}\delta_{p+l-q-r}\,A_pB_lC_qD_r\nonumber\\
=&\,\int_\textbf{p}\int_{l,q,r}\delta_{q+l-p-r}\,A_qB_lC_pD_r\bigg(\frac{p^0}{q^0}\bigg)^{-\phi}\,,
\end{align}
where
\begin{align}
\phi=-\frac{1}{z}\Big(3(d-1)+3z-\alpha-\beta-\gamma-\kappa\Big)
\end{align}
is the exponent of the Jacobian originating from scaling behaviors
\begin{align}
A(p^0,\textbf{p})=&\,s^\alpha A(s^zp^0,s\textbf{p})\,,\,
B(p^0,\textbf{p})=\,s^\beta B(s^zp^0,s\textbf{p})\,,\nonumber\\
C(p^0,\textbf{p})=&\,s^\gamma C(s^zp^0,s\textbf{p})\,,\,
D(p^0,\textbf{p})=\,s^\kappa D(s^zp^0,s\textbf{p})\,.
\end{align}
and the change of variables
\begin{align}
p^0=\frac{p^0}{q^{\prime0}}q^{\prime0}\,,\,\,q^0=\frac{p^0}{q^{\prime0}}p^0\,,\,\,
l^0=\frac{p^0}{q^{\prime0}}l^{\prime0}\,,\,\, 
r^0=\frac{p^0}{q^{\prime0}}r^{\prime0}
\end{align}
and
\begin{align}
\textbf{p}=&\,\bigg(\frac{p^0}{q^{\prime0}}\bigg)^{\frac{1}{z}}\textbf{q}^\prime\,,\,\,\textbf{q}=\bigg(\frac{p^0}{q^{\prime0}}\bigg)^{\frac{1}{z}}\textbf{p}^\prime\,,\,\,\textbf{l}=\bigg(\frac{p^0}{q^{\prime0}}\bigg)^{\frac{1}{z}}\textbf{l}^\prime\,,\nonumber\\
\textbf{r}=&\,\bigg(\frac{p^0}{q^{\prime0}}\bigg)^{\frac{1}{z}}\textbf{r}^{\prime0}.
\end{align}

\bibliography{nonref}

\end{document}